\documentclass[a4paper]{article}

\usepackage{graphicx}

\usepackage{amssymb}

\newcommand{\be}{\begin{equation}}
\newcommand{\ee}{\end{equation}}

\begin{document}


\tolerance=10000

\def\pni{\par \noindent}
\def\vsh{\smallskip}
\def\s{\smallskip}
\def\vs{\medskip}
\def\vvs{\bigskip}
\def\vvvs{\bigskip\medskip} 
\def\vsp{\vsh \pni}
\def\vsn{\vsh\pni}
\def\cen{\centerline}
\def\ra{\item{a)\ }} \def\rb{\item{b)\ }}   \def\rc{\item{c)\ }}
\def\eg{{\it e.g.}\ } 
\hyphenpenalty=2000
\font\bfs=cmbx12 scaled\magstep1
\font\bbfs=cmbx12 scaled \magstep2
\begin{center}
 FRACALMO PRE-PRINT  \  {\tt www.fracalmo.org}
 \\ {\bf 
Journal of Physics A: Math. Theor. Vol. 41 (2008), 285003 (22 p)} 

\vs

\hrule
\vvs

\vvs

{\bbfs{Characterizations and simulations}} 
\vs

{\bbfs {of a class of stochastic processes}}
\\
{\bbfs {to model anomalous diffusion}}
\vvs

Antonio MURA $^{(1)}$ and Gianni PAGNINI$^{(2)}$ 
\vs

$^{(1)}$ Department of Physics, University of Bologna,
 and INFN, \\
   Via Irnerio 46, I-40126 Bologna, Italy\\
   {\tt mura@bo.infn.it}
\vs

$^{(2)}$
National Agency for  New Technologies,
 Energy and the Environment,\\ 
 ENEA, Centre  "E. Clementel", 
 \\ Via Martiri di Monte Sole 4,
 I-40129 Bologna, Italy\\
 {\tt gianni.pagnini@bologna.enea.it}
\vvs

 {\bf Revised Version, June 2008}
\end{center} 
\vvs

\begin{abstract}
In this paper we study a parametric class of stochastic processes to model 
both fast and slow anomalous diffusion. 
This class, called generalized grey Brownian motion (ggBm), 
is made up of self-similar with stationary increments processes 
($H$-{\bf sssi}) and depends on two real parameters 
$\alpha \in (0,2)$ and $\beta \in (0,1]$. 
It includes fractional Brownian motion when $\alpha \in (0,2)$ and 
$\beta =1 $, and time-fractional diffusion stochastic processes when 
$\alpha= \beta \in (0,1)$.
The latters have marginal probability density function governed by 
time-fractional diffusion equations of order $\beta$. 
The ggBm is defined through the explicit construction of the underlying 
probability space. However, in this paper we show that it is possible 
to define it in an unspecified probability space. 
For this purpose, we write down explicitly all the finite dimensional 
probability density functions. 
Moreover, we provide different ggBm characterizations. 
The role of the M-Wright function, which is related to the fundamental 
solution of the time-fractional diffusion equation, emerges as a natural 
generalization of the Gaussian distribution. 
Furthermore, we show that ggBm can be represented in terms of the product 
of a random variable, which is related to the M-Wright function, and an 
independent fractional Brownian motion. 
This representation highlights the $H$-{\bf sssi} nature of the ggBm and 
provides a way to study and simulate the trajectories. 
For this purpose, we developed a random walk model based on a 
finite difference approximation of a partial integro-differenital 
equation of fractional type.
\vvs

\noindent
{\bf PACS numbers: 02.50Ey, 05.40.Fb, 02.30.Rz, 02.30.Gp}

\end{abstract}


\newpage
\section{Introduction}
\label{s1}
\indent

Diffusive processes are generally classified as normal or anomalous if their 
variance grows linearly in time or not, respectively. Furthermore, the normal 
diffusion is associated to the Gaussian probability density function (PDF) for 
particle positions.

Several physical phenomena show anomalous diffusion.
They range from dispersion in complex plasmas
\cite{ratynskaia_etal-prl-2006}
to self-diffusion
of surfactant molecules
\cite{gambin_etal-prl-2005},
or from light in a cold atomic cloud
\cite{labeyrie_etal-prl-2003}
to donor-acceptor electron pair within a protein
\cite{kou_etal-prl-2004},
to mention only some of the more recent experimental evidences.
Such anomalous behaviours cover the full range of
anomalous diffusion, e.g. from slow diffusion
\cite{kou_etal-prl-2004,
solomon_etal-prl-1993,
marty_etal-prl-2005},
when the variance
grows slower than linear, to
fast diffusion
\cite{ratynskaia_etal-prl-2006,
gambin_etal-prl-2005,
labeyrie_etal-prl-2003,
liu_etal-pre-2007},
when the variance grows faster than linear.
In order to model with a unique mathematical framework
both slow and fast anomalous diffusion,
a class of stochastic processes
is here introduced and analyzed. We would like this work to be a first step
towards a comprehensive description of all dispersive mechanisms.
Moreover, the PDF of particle positions
of this class turns out to be related to the $M-$Wright function, which
is a natural generalization of the Gaussian density.

A Gaussian anomalous diffusion can be obtained from a standard diffusion 
equation with time dependent diffusivity. The latter is mainly based on 
the empirical flux-gradient relation and, for this reason, it is considered 
a simple particular case.

Anomalous diffusion processes can also be obtained as Gaussian processes 
with time subordination (see Remark 1). As a consequence, the particle density is not 
Gaussian. This fact is often seen as the origins and the physical 
interpretation of anomalous diffusion. In fact, let $f(x,t)$ be the density 
function of a diffusive process and $G(x,t)$ a standard Gaussian density 
function. Namely, with $t \ge 0$,
\begin{equation}
G(x,t)=\frac{1}{\sqrt{4\pi t}}\exp\left(-\frac{x^2}{4t}\right) \,, \quad
\sigma^2_G(t) :=\int_{\mathbb{R}}x^{2}G(x,t) \, dx =2t \,.
\end{equation}
Let $\varphi_\beta(\tau,t)=t^{-\beta}\phi(\tau t^{-\beta})$, with $\tau \ge 0$,
$t>0$ and $\beta>0$, be the  marginal probability density function of a
self-similar stochastic process, which is interpreted as a randomized operational time. 
Therefore, in agreement with the monotonic growing of time, such a process is required to be 
a non-negative non-decreasing random process. 
Hereinafter, it is called marginal probability density function the one-dimensional PDF of a 
certain stochastic process (e.g.~the one-point and one-time PDF of particle position)\footnote{
This notation is a physical analogue of probability theory notation. In fact, this marginal density corresponds 
to the result of the integration over an $n$-dimensional joint density (e.g. the $n$-points and 
$n$-times density associated to an $n$-steps
particle trajectory)}. Furthermore, we remember that a process $X(t)$, $t\ge 0$,
is said to be self-similar with self-similarity exponent $H$ if, 
for all $a\ge 0$, the processes $X(at)$, $t\ge 0$, and $a^{H}X(t)$, $t\ge 0$,
have the same finite-dimensional distributions. We also suppose that $\phi$ 
has moments of any order. Then, if the subordination formula
\begin{equation}
f(x,t)=\int_0^{+\infty} G(x,\tau) \varphi_\beta(\tau,t) \, d\tau
\label{subordinazione}
\end{equation}
holds, $f(x,t)$ is the marginal density function of an anomalous 
diffusion process. In fact,
\begin{eqnarray*}
\sigma^2_f(t)&=&\int_{-\infty}^{+\infty}x^{2}f(x,t) \, dx
=\int_{-\infty}^{+\infty}x^{2}\left\{
\int_0^{+\infty} G(x,\tau) \varphi_\beta(\tau,t) \, d\tau \right\} \, dx \\
&=&\int_0^{+\infty}\left\{
\int_{-\infty}^{+\infty} x^{2} G(x,\tau) \, dx \right\}
\, \varphi_\beta(\tau,t) \, d\tau\\
&=&2 \int_0^{+\infty}
\tau \, \varphi_\beta(\tau,t) \, d\tau = Dt^{\beta} \,,
\end{eqnarray*}
where we set
$$
D=2\int_0^{+\infty}\zeta\phi(\zeta) \, d \zeta \,,
$$
which is finite by hypothesis.\\

We observe that is desirable having a random time process that for 
$\beta=1$ gives a Gaussian process with a linear growing variance in time. 
Thus, from  \ref{subordinazione}, we require that 
$\varphi_1(\tau,t)=\delta(\tau-t)$.

A ready-made example is given by the $M$-function (see the appendix), 
which is related to the fundamental solution of the so called 
{\it time-fractional diffusion equation} of order $\beta$ 
(see \cite{GorMaiCISM97,Mainardi-Luchko-PagniniFCAA01,PodlubnyBOOK99,
SchneiderFD90,SchneiderWyssJMP89}) which, in integral form, reads:
\begin{equation}
u(x,t)=u_0(x)+\frac{1}{\Gamma(\beta)}\int_{0}^{t}
(t-s)^{\beta-1}\partial_{xx}u(x,s) \,ds \,, \quad t\ge 0 \,,
\label{fraceq}
\end{equation}
with $u_0(x)=u(x,0)$.

In fact, suppose that the marginal density function of the random time 
process is
$$
\varphi(\tau,t)=t^{-\beta}M_{\beta}(\tau t^{-\beta})
\equiv\mathcal{M}_{\beta}(\tau,t) \,,
$$
with $0<\beta<1$. With this choice, \ref{subordinazione} becomes
$$
f(x,t)=\int_0^{+\infty} G(x,\tau) \mathcal{M}_\beta(\tau,t) \, d \tau \,,
$$
which, by using (\ref{GM}) and (\ref{MW}), becomes,
\begin{equation}
f(x,t)=\frac{1}{2}\int_0^{+\infty} \mathcal{M}_{1/2}(|x|,\tau) 
\mathcal{M}_\beta(\tau,t) \, d \tau=
\frac{1}{2} \, \mathcal{M}_{\beta/2}(|x|,t) \,.
\label{subordinazione2}
\end{equation}
That is just the fundamental solution of (\ref{fraceq}). 
Moreover, when $\beta=1$, \ref{fraceq} becomes a standard diffusion 
equation and indeed, using (\ref{pro2}), one finds
$$
f(x,t)=\int_0^{+\infty} G(x,\tau) \mathcal{M}_1(\tau,t) \, d \tau=G(x,t) \,.
$$

\vvs \noindent
 {\bf Example 1.}
Consider a Brownian motion $B(t)$, $t\ge 0$, such that $E(B(1)^2)=2$ 
(we call it standard Brownian motion), and consider a random time 
$l_{\beta}(t)$, $t\ge 0$, defined by the local time\footnote{Heuristically, the local time $l(t,x)$ of a diffusion process characterizes the ``time spent by the process at a given level $x$ up to time $t$'' (we shortly write $l(t)$ if $x=0$). For instance, in the case of Brownian motion, the local time can be written as $l(t,x)=\displaystyle\lim_{\epsilon \rightarrow 0}\frac{1}{2\epsilon}\int_0^t 1_{[x-\epsilon,x+\epsilon]}(B(s))ds$, where $1_{[a,b]}(x)$ is the indicator function of the inteval.} in zero at time $t$ of a 
$d=2(1-\beta)$-dimensional Bessel process, with $0<d<2$ 
(see \cite{Revuz-YorBOOK94,Molchanov-Ostrovskii69} 
and \cite{MainardiCSF96, MainardiCISM97}). 
Furthermore, let $l_\beta$ be independent of $B$. 
It is know that $l_\beta$ is a self-similar process with scaling parameter 
$H=\beta$. Therefore, the subordinated process
\begin{equation}
D_{\beta}(t)=B(l_{\beta}(t)) \,, \quad t\ge 0 \,,
\end{equation}
is a model for slow anomalous diffusion and its marginal probability density 
function is the fundamental solutions of the time fractional diffusion
equation of order $0<\beta\le 1$ (see also \cite{Kolsrud90}).
Actually, the local time $l_\beta(t)$ is defined up to a multiplicative 
constant  (see \cite{donati}). Here, we suppose that $l_{\beta}(t)$, 
$t\ge 0$, is defined such that its marginal density function is  
$\mathcal{M}_{\beta}(x,t)$, $x,t\ge 0$.

\vvs
\noindent
 {\bf Example 2.}
Consider again a standard Brownian motion $B(t)$, $t\ge 0$. Another possible 
choice of an independent random time process $l_\beta(t)$,
for which the subordinated process $B(l_\beta(t))$ has still marginal 
density governed by the time-fractional diffusion equation of
order $\beta$, is the inverse of the totally skewed strictly 
$\beta$-stable process,
as founded in the context of Continuous Time Random Walk (CTRW) 
by  Meerschaert et al.  \cite{meerschaert}
(see also \cite{Metzler-KlafterPhysRep00,BarkaiChemPhys02,GorMaiHONNEF06,stani,scalas}).

\vvs \noindent
{\bf Remark 1.}
The stochastic interpretation through subordinated processes turns out to be 
very natural. A subordinated process is defined as $Y(t)=X(l(t))$, 
$t \ge 0$, where $X(t)$ is a Markovian diffusion and $l(t)$ is a (non-negative) 
random time process independent of $X(t)$ (see \cite{Bochner55,Bochner62}). 
So that, $Y(t)$ has a direct physical interpretation. For instance, $X(t)$ 
can be interpreted as the state of a system at time $t$, while 
$l(t)$ can be interpreted as the ``effective activity'' up to time $t$. 
In this way, even if the process $X(t)$ is Markovian, the resulting 
subordinated process $Y(t)$ is in general non-Markovian and the non-local 
memory effects are attributable to the random time process $l(t)$ and to 
its evolution, which is in general non-local in time.

Examples 1 and 2  provide two different stochastic processes 
with the same marginal density function (\ref{subordinazione2}). 
Indeed, it is important to remark that, starting from a master equation 
which describes the time evolution of a probability density function 
$f(x,t)$, it is always possible to define an equivalence class of stochastic 
processes with the same marginal density function $f(x,t)$. The two examples 
above represent only particular cases of subordinated type processes. 
However, also processes which are not of subordinated type can serve as 
models for anomalous diffusion described by time-fractional diffusion 
equations (see \cite{Mura1}).
It is clear that additional requirements may be stated in order to fix the 
probabilistic model.

Because $0<\beta<1$, \ref{fraceq} depicts a system with a 
slow-anomalous diffusion behavior. In order to study the full range 
(slow and fast) of anomalous diffusion, we introduce a suitable 
{\it time-stretching} $g(t)=t^{\alpha/\beta}$, $0<\beta\le 1$ and $\alpha>0$. 
Let $f(x,t)$ be a solution of (\ref{fraceq}), then the 
function $f_{\alpha,\beta}(x,t)=f(x,t^{\alpha/\beta})$ is a solution of 
the {\it stretched time-fractional diffusion equation}
\begin{equation}
u(x,t)=u_0(x)+\frac{1}{\Gamma(\beta)}\frac{\alpha}{\beta}
\int_{0}^ts^{\frac{\alpha}{\beta}-1}
\left(t^{\frac{\alpha}{\beta}}-s^{\frac{\alpha}{\beta}}\right)^{\beta-1}
\frac{\partial^2}{\partial x^2}u(x,s) \, ds \,,
\label{geneq}
\end{equation}
with the same initial condition. 
Then, the fundamental solution of \ref{geneq} is
$u(x,t)=\frac{1}{2}\mathcal{M}_{\beta/2}(|x|,t^{\alpha/\beta})$ and defines 
a self-similar PDF of order $H=\alpha/2$. That is,
\begin{equation}
u(x,t)=\frac{t^{-\alpha/2}}{2} M_{\beta/2}(|x|t^{-\alpha/2}) \,, 
\quad x\in \mathbb{R} \,.
\label{fonsol}
\end{equation}
The diffusion is slow when $\alpha<1$, standard when 
$\alpha=1$ and fast when $\alpha>1$.
We observe that when
$\beta=1$, $u(x,t)$ is a ``stretched'' Gaussian density
\begin{equation}
u(x,t)=\frac{t^{-\alpha/2}}{2} M_{1/2}(|x|/t^{\alpha/2})=
\frac{1}{\sqrt{4\pi t^{\alpha}}} \exp\left(-\frac{x^2}{4t^\alpha}\right) \,,
\quad t> 0 \,.
\end{equation}
Moreover, in the case $\alpha=\beta$, $0<\beta< 1$, the non-Gaussian 
probability density $u(x,t)=\frac{1}{2}\mathcal{M}_{\beta/2}(|x|,t)$ 
is recovered, i.e. the fundamental solutions of \ref{fraceq}. The diffusion 
is always slow and becomes standard when $\beta\rightarrow 1$. 
Finally, in the general case $0<\beta<1$ and $\alpha>0$, we have a 
non-Gaussian full-ranged anomalous diffusion.\\

Our main goal is to develop stochastic processes that serve as models for 
the anomalous diffusion described by this class of equations 
(\ref{geneq}). To do this we require some constraints.
Let $X(t)$, $t \ge 0$, be a self-similar stochastic process with scaling 
parameter $H=\alpha/2$ and marginal probability density function defined 
by \ref{fonsol}. We have already observed that there is a whole equivalence 
class of such a stochastic processes. For instance, looking at 
Examples 1, 2, one could immediately take 
$X(t)=B(l_\beta(t^{\alpha/\beta}))$, $t\ge 0$, with a suitable choice of the 
independent random time $l_\beta(t)$. In order to choose a specific model, 
we add the requirement that the process $X(t)$ be also a stationary 
increments process. Namely, we require that the process $X(t)$ be 
$H$-{\bf sssi} (self-similar of order $H$ with stationary increments process),
with $H=\alpha/2$.

\vvs\noindent
{\bf Remark 2}
The latter requirement forces the $\alpha$-parameter to be in the range $0<\alpha<2$ 
\cite{TaqquREVBOOK03}. Moreover, it automatically excludes the
subordinated processes of Examples 1, 2, 
which have in general not stationary increments.
Summarizing, we ask that the stochastic process $X(t)$, $t\ge 0$, satisfies the following requirements: let $0<\beta\le 1$ and $0<\alpha<2$, then
\begin{enumerate}
\item[i.] $X(t)$ is self-similar with index $H=\alpha/2$.
\item[ii.] $X(t)$ has marginal density function $f_{\alpha,\beta}(x,t)=
\displaystyle\frac{t^{-\alpha/2}}{2}M_{\beta/2}(|x|t^{-\alpha/2})$
(see (\ref{fonsol})).
\item[iii.] $X(t)$ has stationary increments.
\end{enumerate}
In \cite{Mura1} the Authors shown that a stochastic process which satisfies all 
the above properties is the so called {\it generalized grey Brownian motion} (ggBm) 
$B_{\alpha,\beta}(t)$, $t\ge 0$, \cite{SchneiderGN90a,SchneiderGN90b}. 
It represents a generalization of Brownian motion (Bm) and fractional Brownian
motion (fBm) as well. Moreover, it serves as stochastic model for \ref{geneq}. 
So that, in this paper, we will focus on the study of this process.
\vs \noindent
{\bf Remark 3.}
Because of the stationarity of the increments, 
the anomalous-diffusion appears deeply related to the long-range dependence characterization of
$B_{\alpha,\beta}(t)$. We remember that an $H$-{\bf sssi} process has {\it long-range dependence} 
(or {\it long-memory}) if $1/2<H<1$. This means that
the discrete time process of its increments exhibits long-range correlation. That is, 
the increments autocorrelation function $\gamma(k)$ tends
to zero with a power law as $k$ goes to infinity and in such a manner that it does not 
result integrable 
\cite{TaqquREVBOOK03,MandelbrotJH76,Mandelbrot-Ness68}. 
Therefore, when $0<\alpha<1$ the diffusion is {\it slow} and the process 
has {\it short-memory}. While, when $1<\alpha<2$ the diffusion is 
{\it fast} and the process has {\it long-memory}.

\vs

The rest of the paper is organized as follows: 
in the next section we briefly introduce the mathematical definition of generalized grey Brownian
motion. In the third section we characterize the ggBm through the study of its finite-dimensional 
probability density functions. While, the last two
sections are devoted to trajectory simulations and final remarks.

\section{The generalized grey Brownian motion}
The generalized grey noise space is the probability space $(\mathcal{S}'(\mathbb{R}),\mathcal{B},\mu_{\alpha,\beta})$, where $\mathcal{S}'(\mathbb
{R})$ is the space of tempered distribution defined on $\mathbb{R}$, $\mathcal{B}$ is the Borel's $\sigma$-algebra generated by the weak topology
on $\mathcal{S}'(\mathbb{R})$ and $\mu_{\alpha,\beta}$ is the so called generalized grey noise measure. 
The measure $\mu_{\alpha,\beta}$  satisfies
\begin{equation}
\int_{\mathcal{S}'(\mathbb{R})} e^{i\langle \omega, \xi\rangle}d\mu_{\alpha,\beta}(\omega)=
E_\beta(-||\xi||_{\alpha}^2) \,, \quad \xi \in \mathcal{S}(\mathbb{R}) \,,
\quad 0<\beta\le 1 \,, \quad 0<\alpha<2 \,,
\label{gnm}
\end{equation}
where $\langle \cdot, \cdot \rangle$ is the canonical bilinear pairing between 
$\mathcal{S}(\mathbb{R})$ and $\mathcal{S}'(\mathbb{R})$ (see \cite{Mura1})
and $E_{\beta}(t)$ is the Mittag-Leffler function of order $\beta$
\begin{equation}
E_\beta(x)=\sum_{n=0}^{\infty}\frac{x^n}{\Gamma(\beta n+1)} \,, \quad x\in \mathbb{R} \,.
\label{equ1}
\end{equation}
Moreover,
\begin{equation}
||f||_{\alpha}^2=\Gamma(1+\alpha)\sin{\frac{\pi}{2}\alpha}\int_{\mathbb{R}}
dx |x|^{1-\alpha}|\widetilde f(x)|^2 \,, \quad f\in \mathcal{S}(\mathbb{R}) \,,
\end{equation}
with
$$
\widetilde f(x)=\frac{1}{\sqrt{2\pi}}\int_{\mathbb{R}}dy e^{ixy}f(y) \,,
\quad f\in \mathcal{S}(\mathbb{R}) \,.
$$
The range $0<\beta\le 1$ ensures the complete monotonicity of the function 
$E_{\beta}(-t)$, $t\ge 0$ (see W R Schneider \cite{SchneiderCM-ML96}), 
as required by \ref{gnm}, while the range $0<\alpha<2$ is chosen in 
order to have $||1_{[a,b)}||_\alpha^2<\infty$, where $1_{[a,b)}$ is
the indicator function of the interval $[a,b)$. 
In fact, in this case one has
\begin{equation}
||1_{[a,b)}||_\alpha^2=(b-a)^\alpha \,, \quad 0<\alpha<2 \,, \quad 0\le a<b \,.
\label{indc}
\end{equation}
It is possible to show (see \cite{Mura1,SchneiderGN90b, SchneiderGN90a}) 
that for each $t>0$ the real random variable
\begin{equation}
X_{\alpha,\beta}(1_{[0,t)})(\cdot)=\langle \cdot, 1_{[0,t)}\rangle
\label{gbm1}
\end{equation}
is defined almost everywhere on $\mathcal{S}'(\mathbb{R})$. 
Moreover, it follows from (\ref{gnm}) that it belongs to 
$L^2(\mathcal{S}'(\mathbb{R}),\mu_{\alpha,\beta})$ and
$$
E(X_{\alpha,\beta}(1_{[0,t)})^2)=\frac{2}{\Gamma(1+\beta)}t^{\alpha} \,.
$$
The generalized grey Brownian motion is then defined as the process
\begin{equation}
B_{\alpha,\beta}(t)=X_{\alpha,\beta}(1_{[0,t)}),\;\; t\ge 0.
\label{ggbmdef}
\end{equation}
The $B_{\alpha,\beta}(t)$ marginal density function, indicated with 
$f_{\alpha,\beta}(x,t)$, is the fundamental solution of \ref{geneq}. 
Namely, 
$f_{\alpha,\beta}(x,t)=\frac{t^{-\alpha/2}}{2}M_{\beta/2}(|x|t^{-\alpha/2})$ 
(see Remark 5). Moreover, the linearity of definition (\ref{ggbmdef})
can be used to show many of the fundamentals properties of 
$B_{\alpha,\beta}(t)$. For instance, $B_{\alpha,\beta}(t)$ turns out to be 
$H$-{\bf sssi} with $H=\alpha/2$. Furthermore, one can calculate 
characteristic functions. For example in the one dimensional case, 
for any real $y$ and $t>0$,
$$
\widetilde f_{\alpha,\beta}(y,t)=E\left(e^{iy(B_{\alpha,\beta}(t))}\right)=
E\left(e^{iX_{\alpha,\beta}\left(y1_{[0,t)}\right)}\right) \,.
$$
By using Equations (\ref{gnm}) and (\ref{indc}), one has
$$
\widetilde f_{\alpha,\beta}(y,t)=E_{\beta}(-y^2||1_{[0,t)}||^2_\alpha)=
E_\beta(-y^2t^\alpha) \,.
$$
In the multidimensional case, given a sequence of real numbers 
$\{\theta_{1},\theta_2,\dots, \theta_n\}$,  for any collection 
$\{B_{\alpha,\beta}(t_1),\dots,B_{\alpha,\beta}(t_n)\}$ with $0<t_1<t_2<\cdots <t_n$, 
using linearity again, one can show that \cite{Mura1}
\begin{equation}
E\left(\exp\Big{(}i\sum_{j=1}^{n}\theta_jB_{\alpha,\beta}(t_j)\Big{)}\right)=
E_{\beta}\left(-\Gamma(1+\beta)\frac{1}{2}\sum_{i,j=1}^{n}\theta_i\theta_j
\gamma_{\alpha,\beta}(t_i,t_j)\right) \,,
\label{furloc}
\end{equation}
where
\begin{equation}
\gamma_{\alpha,\beta}(t,s)=\frac{1}{\Gamma(1+\beta)}\left(t^\alpha+s^\alpha-|t-s|^\alpha\right) \,,
\quad t,s \ge 0 \,,
\label{gbmcov}
\end{equation}
is the autocovariance matrix of $B_{\alpha,\beta}(t)$.

\vs \noindent
{\bf Remark 4}
From \ref{furloc}, it follows that, with $\beta$ fixed, $B_{\alpha,\beta}(t)$ 
is defined only by its covariance structure.
In other words, the ggBm, which is not Gaussian in general, is an example of a process defined only through its first and second moments, which is a property of Gaussian processes indeed.

\section{Characterization of the ggBm}
We want now to characterize the ggBm through its finite dimensional structure.
From \ref{furloc}, we know that all the ggBm finite dimensional
probability density functions are defined only by their autocovariance matrix.
The following proposition holds
\newtheorem{Proposition}{Proposition}
\begin{Proposition}
$\null$ \\
Let $B_{\alpha,\beta}$ be a ggBm, then for any collection 
$ \{B_{\alpha,\beta}(t_1),\dots, B_{\alpha,\beta}(t_n)\}$, 
the joint probability density function is given by
\begin{equation}
f_{\alpha,\beta}(x_1,\dots,x_n;\gamma_{\alpha,\beta})=
\frac{(2\pi)^{-\frac{n-1}{2}}}
{\sqrt{2\Gamma(1+\beta)^n\det{\gamma_{\alpha,\beta}}}} 
\int_{0}^{\infty}\frac{1}{\tau^{n/2}}M_{1/2}\left(\frac{\xi}{\tau^{1/2}}\right)
M_\beta(\tau) \, d\tau \,,
\label{mdim}
\end{equation}
with
$$
\xi=\left(2\Gamma(1+\beta)^{-1}\sum_{i,j=1}^{n}x_i
{\gamma_{\alpha,\beta}}^{-1}(t_i,t_j)x_j\right)^{1/2} \,,
$$
$$
\gamma_{\alpha,\beta}(t_i,t_j)=\frac{1}{\Gamma(1+\beta)}
(t_i^{\alpha}+t_j^{\alpha}-|t_i-t_j|^{\alpha}) \,, \quad i,j=1,\dots,n \,.
$$
\end{Proposition}
\noindent
{\bf Proof}: in order to show \ref{mdim}, we calculate its $n$-dimensional 
Fourier transform and we find that it is equal to (\ref{furloc}). We have
$$
\int_{\mathbb{R}^n}\exp\left(i\displaystyle\sum_{j=1}^{n}\theta_jx_j\right)
f_{\alpha,\beta}(x_1,\dots,x_n;\gamma_{\alpha,\beta}) \, d^nx=
\qquad \qquad \qquad \qquad
\qquad \qquad \qquad \qquad
$$
$$
\frac{(2\pi)^{-\frac{n-1}{2}}}
{\sqrt{2\Gamma(1+\beta)^n\det{\gamma_{\alpha,\beta}}}}
\int_{0}^{\infty}\frac{1}{\tau^{n/2}}M_\beta(\tau)
\int_{\mathbb{R}^n}\exp\left(i\displaystyle\sum_{j=1}^{n}\theta_jx_j\right)
M_{1/2}\left(\frac{\xi}{\tau^{1/2}}\right) d^n x d\tau \,.
$$
We remember that $M_{1/2}(r)=\frac{1}{\sqrt \pi}e^{-r^2/4}$, thus we get
$$
\int_{0}^{\infty}\frac{1}{\tau^{n/2}}M_\beta(\tau)\int_{\mathbb{R}^n}
\exp\left(i\displaystyle\sum_{j=1}^{n}\theta_jx_j\right) \times
\qquad \qquad \qquad \qquad
\qquad \qquad \qquad \qquad
\qquad \qquad \qquad \qquad
$$
\begin{equation}
\qquad
\frac{(2\pi)^{-\frac{n}{2}}}{\sqrt{\Gamma(1+\beta)^n\det{\gamma_{\alpha,\beta}}}}
\exp\left(-\sum_{i,j=1}^{n}
\frac{x_i\gamma_{\alpha,\beta}^{-1}(t_i,t_j)x_j}{2\tau\Gamma(1+\beta)}\right)
\, d^nxd\tau \,.
\label{fact0}
\end{equation}
We make the change of variables ${\bf{x}}=\Gamma(1+\beta)^{1/2}\tau^{1/2} {\bf{y}}$, 
with ${\bf{x,y}}\in \mathbb{R}^n$, and we get
$$
\int_{0}^{\infty}M_\beta(\tau)\int_{\mathbb{R}^n}\exp\left(i\Gamma(1+\beta)^{1/2}\tau^{1/2}
\sum_{j=1}^{n}\theta_jy_j\right) \times 
\qquad \qquad \qquad \qquad
\qquad \qquad
$$
$$
\qquad \qquad \qquad \qquad
\qquad \qquad
\frac{(2\pi)^{-\frac{n}{2}}}{\sqrt{\det{\gamma_{\alpha,\beta}}}}\exp\left(-\sum_{i,j=1}^{n}
\frac{y_i\gamma_{\alpha,\beta}^{-1}(t_i,t_j)y_j}{2}\right) \, d^ny d\tau =
$$
$$
\int_{0}^{\infty}M_\beta(\tau)\exp\left(-\Gamma(1+\beta)\tau
\sum_{i,j=1}^{n}
\frac{\theta_i \gamma_\alpha(t_i,t_j) \theta_j}{2}\right)d\tau=
\int_{0}^{\infty}e^{-\tau s}M_{\beta}(\tau)d\tau=E_{\beta}(-s) \,,
$$
where
$s=\Gamma(1+\beta)\sum_{i,j=1}^{n}\theta_i\theta_j\gamma_{\alpha,\beta}(t_i,t_j)/2$
and we have used \ref{LaM}. $\Box$\\

Applying the Kolmogorov extension theorem,
the above proposition allows us to define the ggBm in an unspecified probability space. 
In fact, given a probability space $(\Omega,\mathcal{F},P)$, the following proposition characterizes 
the ggBm
\begin{Proposition}
\label{pp1}
Let $X(t)$, $t\ge 0$, be a stochastic process defined in a certain probability space 
$(\Omega,\mathcal{F},P)$, such that
\begin{enumerate}
\item[i)] $X(t)$ has covariance matrix indicated by $\gamma_{\alpha,\beta}$ and 
finite-dimensional distributions defined by \ref{mdim},
\item[ii)] $E(X^2(t))=\frac{2}{\Gamma(1+\beta)}t^{\alpha}$
for $0<\beta\le 1$ and $0<\alpha<2$,
\item[iii)] $X(t)$ has stationary increments,
\end{enumerate}
then $X(t)$, $t\ge 0$, is a generalized grey Brownian motion.
\end{Proposition}
In fact condition ii) together with condition iii) imply that $\gamma_{\alpha,\beta}$ must be the 
ggBm autocovariance matrix (\ref{gbmcov}).

\vs \noindent
{\bf Remark 5}
Using (\ref{MW}), for $n=1$, \ref{mdim} reduces to
\begin{eqnarray}
f_{\alpha,\beta}(x,t) &=&
\displaystyle\frac{1}{\sqrt{4t^{\alpha}}}\int_{0}^{\infty}
\mathcal{M}_{1/2}\left(|x|t^{-\alpha/2},\tau\right)
\mathcal{M}_\beta(\tau,1) \, d\tau \nonumber \\
&=&\frac{1}{2}t^{-\alpha/2}M_{\beta/2}(|x|t^{-\alpha/2}) \,.
\label{eqone}
\end{eqnarray}
This means that the ggBm marginal density function is indeed the fundamental solution of 
\ref{geneq}.

\vs \noindent
{\bf Remark 6}
Because for $\beta=1$
$$
M_1(\tau)=\delta(\tau-1) \,,
$$
then, putting $\gamma_{\alpha,1}\equiv \gamma_\alpha$,
we have that \ref{mdim} reduces to the Gaussian distribution of the
fractional Brownian motion. That is,
$$
f_{\alpha,1}(x_1,x_2,\dots,x_n;\gamma_{\alpha,1})=\displaystyle\frac{(2\pi)^{-\frac{n-1}{2}}}{\sqrt{2\det{\gamma_{\alpha}}}}M_{1/2}\left(\left(2\sum_{i,j=1}^{n}x_i\gamma_{\alpha}^{-1}(t_i,t_j)x_j\right)^{1/2}\right).
$$
We have the following corollary.
\newtheorem{Corollary}{Corollary}
\begin{Corollary}
Let $X(t)$, $t\ge 0$, be a stochastic process defined in a certain probability space
$(\Omega,\mathcal{F},P)$. Let $H=\alpha/2$ with $0<\alpha<2$ and suppose that
$E(X(1)^2)=2/\Gamma(1+\beta)$. The following statements are equivalent
\begin{itemize}
\item[i)] $X$ is $H$-{\bf sssi}
with finite-dimensional distribution defined by (\ref{mdim}),
\item[ii)] $X$ is a generalized grey Brownian motion with scaling exponent $\alpha/2$ and ``fractional order'' parameter $\beta$,
\item[iii)] $X$ has zero mean, covariance function $\gamma_{\alpha,\beta}(t,s)$, $t,s\ge 0$, 
defined by (\ref{gbmcov}) and finite dimensional distribution defined by (\ref{mdim}).
\end{itemize}
\end{Corollary}
\subsection{Representation of ggBm}
Up to now, we have seen that the ggBm $B_{\alpha,\beta}(t)$, $t\ge 0$, is an $H$-{\bf sssi} process, which generalizes Gaussian processes (it is indeed Gaussian when $\beta=1$) and is defined only by its autocovariance structure. These properties make us think that $B_{\alpha,\beta}(t)$ may be equivalent to a process $\Lambda_\beta X_\alpha(t)$, $t\ge 0$, where $X_{\alpha}(t)$ is a Gaussian process and $\Lambda_\beta$ is a suitable chosen independent random variable. Indeed, the following proposition holds
\begin{Proposition}
\label{pp2}
Let $B_{\alpha,\beta}(t)$, $t\ge 0$, be a ggBm, then
\begin{equation}
B_{\alpha,\beta}(t)=^{\!\!\!\!{}^{d}}
\sqrt{L_\beta}X_\alpha(t) \,, \quad t\ge 0 \,, \quad 0<\beta\le 1 \,, \quad 0<\alpha<2 \,,
\label{ggrep}
\end{equation}
where $=^{\!\!\!\!{}^{d}}$ denotes the equality of the finite dimensional distribution, 
$X_\alpha(t)$ is a standard fBm and $L_\beta$ is an independent non-negative random variable 
with probability density function $M_\beta(\tau)$, $\tau\ge 0$.
\end{Proposition}
In fact, after some manipulation, \ref{fact0} can be written as follows
$$
\int_{\mathbb{R}^n}\int_{0}^{\infty}
\exp\left(i\displaystyle\sum_{j=1}^{n}\theta_jyx_j\right)2yM_\beta(y^2)
\displaystyle\frac{(2\pi)^{-\frac{n}{2}}}{\sqrt{\det{\gamma_\alpha}}}
\exp\left(-\sum_{i,j=1}^{n}x_i\gamma_\alpha^{-1}(t_i,t_j)x_j/2\right) dyd^nx =
$$
$$
\qquad \qquad \qquad \qquad
E\left(\exp(i \sum_{j=1}^{n}\theta_j\sqrt{L_\beta} X_{\alpha}(t_j))\right) \,.
$$
\vs\noindent
{\bf Example 3}
A possible choice of $L_\beta$ is the random variable
$l_\beta(1)$, where $l_\beta(t)$, $t\ge 0$, is the random time process of
Example 1 or Example 2.

\vs\noindent
{\bf Remark 7}
Proposition \ref{pp2} highlights the $H$-{\bf sssi} nature of the ggBm. Moreover,
for $\beta=1$ from \ref{pro2} follows that $L_1=1$ a.s.,
thus we recover the fractional Brownian motion of order $H=\alpha/2$.

\vs\noindent
Representation (\ref{ggrep}) is very interesting.
In fact, a number of question, in particularly those related to the distribution properties
of $B_{\alpha,\beta}(t)$, can be reduced to questions concerning the fBm $X_\alpha(t)$,
which are easier since $X_\alpha(t)$ is a Gaussian process.
For instance, the H\" older continuity of the  $B_{\alpha,\beta}(t)$ trajectories follows
immediately from those of $X_\alpha(t)$
$$
E(|X_\alpha(t)-X_{\alpha}(s)|^p)=c_p|t-s|^{p\alpha/2} \,.
$$
Moreover, this factorization is indeed suitable for path-simulation (see next section).

\vs \noindent
{\bf Remark 8}
From (\ref{ggrep}), it is clear that the Brownian motion
($B_{1,1}(t)$, $t\ge 0$) is the only one process of the ggBm class with
independent increments.

\section{Path simulation}
In the previous section we have shown that the ggBm could be represented by
the process
$$
B_{\alpha,\beta}(t)=\sqrt{L_{\beta}}X_\alpha(t),\;\;t\ge 0\,, \quad
0 < \beta \le 1\,, \quad 0 < \alpha < 2 \,,
$$
where $L_\beta$ is a suitable chosen random variable independent of the standard fBm $X_\alpha(t)$.
Clearly, to simulate ggBm trajectories we first need a method to generate
the random variable $L_\beta$.

\subsection{The time fractional drift equation}
In order to generate the random variable $L_\beta$ with probability density function
$M_\beta(\tau)$,
we consider the so called {\it time-fractional forward drift equation},
which in integral form reads
\begin{equation}
u(x,t)=u_0(x)-\frac{1}{\Gamma(\beta)}\int_{0}^{t}(t-s)^{\beta-1}\frac{\partial}{\partial x}u(x,s) \, ds \,,
\quad x\in \mathbb{R} \,, \quad t\ge 0 \,, \quad 0<\beta\le 1\,.
\label{tfdrift}
\end{equation}
The fundamental solution of \ref{tfdrift} is 
\cite{MainardiCSF96,MainardiCISM97}
\begin{equation}
u(x,t)=\mathcal{M}_{\beta}(x,t) \,, \quad x,t \ge 0 \,.
\label{lfund1}
\end{equation}
This function can be interpreted as the marginal density function of a non-negative
self-similar stochastic process with scaling parameter $H=\beta$
(see Examples 1, 2).

\vs \noindent
{\bf Remark 9}
The name {\it ``drift equation''} refers to the fact that when 
$\beta=1$ \ref{tfdrift} turns out to be the
one dimensional (forward) drift equation 
$\partial_t u(x,t)=-\partial_x u(x,t)$, whose fundamental solution is  $\delta (x-t)$.

\vs\noindent
{\bf Remark 10}
When $\beta =1$ we recover $\mathcal{M}_{1}(x,t)=\delta(x-t)$  
(see (\ref{pro2})).

\vs
We write (\ref{tfdrift}{} in terms of fractional derivative of order 
$\beta$. Let us introduce the {\it Caputo-Dzherbashyan} derivative
\begin{equation}
_*D^{\beta}_tf(t)=J_t^{1-\beta}\frac{\partial f}{\partial t}=
\frac{1}{\Gamma(1-\beta)}\int_{0}^{t} (t-s)^{-\beta}
\frac{\partial f(s)}{\partial s} \, ds \,,
\label{CD}
\end{equation}
where $J_t^\alpha$ is the Riemann-Liouville fractional integral of order 
$0\le\alpha<1$ such that, for $\alpha=0$, $J_t^0$ is the identity operator. 
Then, the corresponding Cauchy problem of \ref{tfdrift} can be written
\begin{equation}
\left\{
\begin{array}{ll}
_*D_t^{\beta}u(x,t)=-\partial_x u(x,t) \,, \\
u(x,0)=u_0(x)=\delta(x) \,,
 \end{array}
 \right.
\label{cap}
\end{equation}
with $x\in \mathbb{R}$, $t\ge 0$ and $0<\beta\le 1$.

Using a random walk model, one can simulate a discrete time random process
$L_\beta(t)$, $t\ge 0$, governed by the time-fractional forward drift equation (\ref{cap})
(see \cite{GorMaiNLD02,GorMaiCHEMPHYS02}).
In this way, for each run, the random variable $L_\beta(1)$ has the required
distribution $u(x,1)=M_\beta(x)$. The random walk construction follows two steps:
\begin{itemize}
\item the Gr\" unwald-Letnikov discretization of Caputo-Dzherbashyan derivative,
\item the interpretation of the corresponding finite difference scheme as a random walk scheme.
\end{itemize}

\subsection{Finite difference schemes}
In order to define the finite difference model, we write the Cauchy problem (\ref{cap})
in a finite domain
\begin{equation}
\bf
\left\{
\begin{array}{ll}
_*D_t^{\beta}u(x,t)=-\frac{\partial}{\partial x} u(x,t) \,, \quad (x,t)\in \Omega=[-a,a]\times [0,1] \,,
\quad a>0 \,,\\
u(x,0)=u_0(x)=\delta(x) \,,\\
u(-a,t)=\Phi_1(t) \,, \quad u(a,t)=\Phi_2(t) \,, \quad t>0 \,.
\end{array}
\right.
\end{equation}
Let $N,M$ be positive integers. Then, we introduce a bi-dimensional lattice
$$
\mathcal{G}^{2M,N}_{\delta x,\delta t}=
\{(j\delta x,n\delta t),(j,n)\in\mathbb{Z}_{2M}\times \mathbb{Z}_{N}\}\,,
$$
contained on $\Omega$, with $\delta x=2a/(2M-1)$ and $\delta t=1/(N-1)$. 
The lattice elements are indicated with
$$
(x_{j},t_{n})=(j\delta x,n\delta t) \,, \quad j=0,1,\dots, 2M-1 \,, \quad 
n=0,1,\dots N-1 \,.
$$
Let $u:\Omega \rightarrow \mathbb{R}$ be a function defined on 
$\Omega$. We indicate with $u^{n}_j=u(x_j,t_n)$ the restriction of $u$ to 
$\mathcal{G}^{2M,N}_{\delta x,\delta t}$ evaluated in $(x_j,t_n)$.\\

The time-fractional forward drift equation is then replaced by the finite difference equation
\begin{equation}
_*D_{t}^{\beta}u_j^{n}=-\frac{u_{j}^{n}-u_{j-1}^n}{\delta x},
\end{equation}
where
\begin{equation}
\begin{array}{ll}
_*D_{t}^{\beta}u_j^n=\displaystyle\sum_{k=0}^{n+1}(-1)^k
{\beta \choose k}
\frac{u_j^{n+1-k}-u_j^0}{\delta t^{\beta}}\,,
\quad u^0_j=u_0(j\delta x) \,,\\[0.6cm]
\displaystyle{\beta \choose k}=\frac{\Gamma(\beta+1)}{\Gamma(k+1)\Gamma(\beta-k+1)},
\end{array} 
\end{equation}
is the so called forward Gr\" unwald-Letnikov scheme for the Caputo-Dzherbashyan derivative (\ref{CD}).
Using the ``empty sum'' convention
$$
\sum_{k=p}^{q}\cdot =0 \,, \quad \textrm{if }q<p \,,
$$
for any $n \ge 0$, we obtain the explicit equation
\begin{equation}
u_{j}^{n+1}=u_j^0\sum_{k=0}^{n}(-1)^k
\left({\beta}\atop{k}\right)
+\sum_{k=1}^{n}(-1)^{k+1}
\left({\beta}\atop{k}\right) u_j^{n+1-k}+\mu(u_{j-1}^{n}-u_{j}^n),
\label{one}
\end{equation}
where $\mu=\delta t^{\beta}/\delta x$. Equation (\ref{one}) can be written in the following noteworthy form
\begin{equation}
u_{j}^{n+1}=b_nu_j^0+\sum_{k=1}^{n}c_ku_j^{n+1-k}+\mu(u_{j-1}^{n}-u_j^n),
\label{expl}
\end{equation}
where we have defined
\begin{equation}
\begin{array}{ll}
c_k =
(-1)^{k+1}
\displaystyle\left({\beta}\atop{k}\right) \,, \quad k\ge 1 \,,\\
\\
b_n = \sum_{k=0}^{n}(-1)^k
\displaystyle\left({\beta}\atop{k}\right) \,, \quad n\ge 0 \,.
\end{array}
\end{equation}
More precisely, the explicit scheme reads
$$
\left\{
\begin{array}{ll}
u^{n}_{0}=\Phi_1(t_n),\;\; u^{n}_{2M-1}=\Phi_2(t_n),\;\; n> 0,\\[0.3cm]
u_j^1=(1-\mu+\mu L)u^0_j,\;\; 0<j<2M-1\\[0.3cm]
u_{j}^{n+1}=(c_1-\mu +\mu L)u^{n}_j+c_2u_j^{n-1}+\dots +c_nu^{1}_j+b_nu_j^0,\;\; n> 0,\;\; 0< j< 2M-1,
\end{array}
\right.
$$
where $L$ is the ``lowering'' operator  $Lf_j=f_{j-1}$.

\vs\noindent
{\bf Remark 11}
When $\beta=1$ all the coefficients $c_k$ and $b_n$ vanish except $b_0=c_1=1$.
So that, we recover the finite difference approximation of the (forward) drift equation.

\vs
It order to write an implicit scheme, we have to use backward approximations
for the time-fractional derivative. Therefore, for any $n\ge 0$, we obtain
\begin{equation}
u_{j}^{n+1}+\mu(u_{j}^{n+1}-u_{j-1}^{n+1})=b_nu_j^0+\sum_{k=1}^{n}c_ku_j^{n+1-k}.
\label{impl}
\end{equation}
Namely,
$$
\left\{
\begin{array}{ll}
u^{n}_{0}=\Phi_1(t_n),\;\;u^{n}_{2M-1}=\Phi_2(t_n),\;\; n> 0\\
(1+\mu-\mu L)u_j^1=u^0_j,\;\;\;0< j< 2M-1,\\
(1+\mu-\mu L)u_{j}^{n+1}=c_1u^{n}_j+c_2u_j^{n-1}+\dots +c_nu^{1}_j+b_nu_j^0,\; n>0,\;\;0<j<2M-1.
\end{array}
\right.
$$
The above equation can be rewritten in matrix notation
\begin{equation}
\left\{
\begin{array}{ll}
\Lambda_{ij}u_j^1=u_i^0+\psi^1_i,\\[0.3cm]
\Lambda_{ij}u_{j}^{n+1}=c_1u_i^{n}+c_2u_i^{n-1}+\dots +c_nu_i^{1}+b_nu_i^0+\psi^{n+1}_i,\;\; n\ge 1.\\[0.3cm]
\end{array}
\right.
\end{equation}
$\Lambda$ is the following $2M\times 2M$ matrix, divided in four 
$M\times M$ blocks
$$
\Lambda=
\left(
\begin{array}{c|c}
\Lambda_1 & 0\\
\hline
\Lambda_2 & A
\end{array}
\right)
$$

\begin{equation}
\Lambda=
\left(
\begin{array}{ccccc|ccccccccc}
1 & 0 & 0 & \cdots & 0 & 0 & 0 & \cdots & \cdots & 0\\
-\mu & 1+\mu & 0 & \cdots & 0 & 0 & 0 & \cdots & \cdots & 0\\
0 & -\mu & 1+\mu  & \cdots & 0 & 0 & 0 & 0 & \cdots & 0\\
\vdots & \vdots &\vdots & \ddots & \vdots & \vdots & \vdots & \vdots & \ddots & \vdots\\
0 & 0 & \cdots & -\mu & 1+\mu & 0 & 0 & \cdots & 0 & 0\\
\hline
 0 & 0 & 0 & \cdots & -\mu & 1+\mu & 0 & 0 & \cdots & 0\\
 0 & 0 & 0 & \cdots & 0 &-\mu & 1+\mu & 0 & \cdots & 0\\
 0 & 0 & 0 & \cdots & 0 & 0 & -\mu & 1+\mu  & \cdots & 0\\
 \vdots & \vdots & \vdots & \ddots & \vdots & \vdots & \vdots & \vdots & \ddots & \vdots\\
 0 & 0 & 0 & \cdots & 0 & 0 & 0 & \cdots & -\mu & 1
\end{array}
\right)
\,.
\label{A}
\end{equation}
Moreover, for any $n\ge 0$, $\psi^{n}$ is a suitable vector which takes 
into account for the boundary terms.

\vs\noindent
{\bf Remark 12}
Because $\Lambda$ is lower diagonal, $\Lambda^{-1}$ is
\begin{equation}
\Lambda^{-1}=\left(
\begin{array}{c|c}
\Lambda_1^{-1} & 0\\
\hline
-A^{-1}\Lambda_2\Lambda_1^{-1} & A^{-1}
\end{array}
\right)
\,.
\label{lainv}
\end{equation}

\vs
As usual, the explicit scheme is subjected to a stability condition, while te implicit scheme is always stable. For example, if we take $\mu\le\beta$, namely
\begin{equation}
\delta x \ge \delta t^\beta/\beta,
\label{stab}
\end{equation}
then the explicit scheme results indeed stable and preserves non-negativity as well. This means that if we suppose $u^0_j\ge 0$ for any $0\le j\le 2M-1$, then $u^n_j\ge 0$ for any $n>0$ and $0<j<2M-1$. Actually, this is crucial because we interpret the $\{u^n_j\}$ as sojourn probabilities. In order to show this, it is convenient to write equations (\ref{expl}) and (\ref{impl}) in the Fourier domain. Namely, we apply the {\it discrete Fourier transform} with respect to the ``space'' index $j$ to both sides of (\ref{expl}) and (\ref{impl}). We remember that, given a collection of complex numbers $\{x_j,\; j=0,1,\cdots, 2M-1\}$\footnote{Actually, we are considering $\{x_j\}$, for any $j\in \mathbb{Z}_+$, as a periodic sequence, such that $x_{j+2M}=x_j$ for any non-negative integer $j$. }, then its discrete Fourier transform is usually defined as:
$$
\mathcal{F}_d (x_l)_k:=\widehat x_k=\sum_{j=0}^{2M-1}x_je^{-i2\pi j k /2M},\;\; k=0,\cdots, 2M-1.
$$     
One can show that, for any real number $a$, one has:
$$
\mathcal{F} (x_{j+a})_k=e^{i2\pi ak/2M}\widehat x_k.
$$
Thus, heuristically, the effect of applying the Fourier transform to our finite difference equations is just to ``line up'' the points in the Fourier space. In fact, for any $n>0$, we get:
\begin{equation}
\left\{
\begin{array}{ll}
\widehat u^{n+1}_k=\xi_e(k)\widehat u^{n}_k+(\cdots)_k, & \textrm{explicit case},\\[0.3cm]
\xi_i(k)^{-1}\widehat u^{n+1}_k=\beta \widehat u^{n}_k+(\cdots)_k,& \textrm{implicit case},
\end{array}
\right.
\end{equation}
where $\xi_e(k)=(\beta-\mu+\mu e^{-i\pi k/M})$ and $\xi_i(k)=(1+\mu-\mu e^{-i\pi k/M})^{-1}$ correspond to the so called {\it amplification factors}. Then, heuristically, the schemes are stable if $|\,\xi(k)|\le 1$ for any $k$. Namely, one has to require:
\begin{equation}
\left\{
\begin{array}{ll}
\displaystyle\max_k \left((\beta-\mu)^2+\mu^2+2\mu(\beta-\mu)\cos (\pi k/M) \right)\le 1,& \textrm{explicit case},\\[0.3cm]
\displaystyle \min_k \left((1+\mu)^2+\mu^2-2\mu(1+\mu)\cos(\pi k/M)\right)\ge 1, & \textrm{implicit case}.
\end{array}
\right.
\end{equation}
Clearly, while the second one is always satisfied, the first condition is indeed true if (\ref{stab}) 
holds\footnote{This follows from the fact that if $(\beta-\mu)$ is non-negative, the maximum is reached when 
the cosine equals $+1$, then one has $\beta^2\le 1$, which is in fact true by hypothesis.}. 
\subsection{Random walk models}
We observe that, for $0<\beta<1$,
\begin{equation}
\sum_{k=1}^{\infty} c_k=1,\;\; 1>\beta=c_1>c_2>\dots \rightarrow 0 \,.
\label{cp}
\end{equation}
Moreover,
\begin{equation}
\left\{
\begin{array}{ll}
b_0=1=\displaystyle\sum_{k=1}^{\infty}c_k,\;\; b_m=1-\displaystyle\sum_{k=1}^{m}c_k=
\displaystyle\sum_{k=m+1}^{\infty}c_k \,,\\
1=b_0>b_1>b_2>\dots \rightarrow 0 \,.
\end{array}
\right.
\label{bp}
\end{equation}
Thus, the coefficients $c_k$ and $b_n$ are
a sequence of positive numbers, which do not exceed unity and decrease strictly monotonically to zero.

\subsubsection{Explicit random walk}
In order to build a random walk model, we consider for first the explicit scheme. 
Omitting the boundary terms, we have
$$
\left\{
\begin{array}{ll}
u_j^1=(1-\mu)u^0_j+\mu u^0_{j-1} \,,\\
u_{j}^{n+1}=(c_1-\mu)u^{n}_j +\mu u^{n}_{j-1}+c_2u_j^{n-1}+\dots +c_nu^{1}_j+b_nu_j^0 \,,
\quad n\ge 1 \,.
\end{array}
\right.
$$
We consider a walker which starts in zero at time zero, namely $x(t=0)=0$. 
We interpret the $u^n_j$ as the probability of sojourn in $x_j=j\delta x$ at time $t_n=n\delta t$. 
Then, we indicate with $x(t_n)$ the position of the particle at time $t_n$.

At time $t_1=\delta t$ the walker could be at the position $x(1)=x_1$ with probability $\mu$ (that is the probability to come from one space-step behind) or in $x(1)=x_0=0$ with probability $1-\mu$ (that is the probability to remain in the starting position).

From Equations (\ref{cp}) and (\ref{bp}), 
it is clear that the parameters $c_1,c_2,\dots c_n, b_n$ can be interpreted as probabilities.
Then, the position at time $t_{n+1}$ is determined as follows.
We define a partition of events $\{E_{c_1},E_{c_2},\dots, E_{c_n}, E_{b_n}\}$, with $P(E_{c_k})=c_k$, $P(E_{b_n})=b_n$, $n\ge 1$ and such that:
\begin{itemize}
\item $E_{c_1}=$\Big{\{}{\it the particle starts in the previous position $x(t_{n})$ and jumps in $x(t_n)+\delta x$ with probability $\mu$ or stays in $x(t_n)$ with probability $1-\mu$}\Big{\}}.
\item $E_{c_k}=$\Big{\{}{\it the particle backs to the position $x(t_{n+1-k})$}\Big{\}}.
\item $E_{b_n}=$\Big{\{}{\it the particle backs to the initial position $x(t_{0})$}\Big{\}}.
\end{itemize}
\subsubsection{Implicit random walk}
Consider the implicit case
$$
\left\{
\begin{array}{ll}
u^1=\Lambda^{-1}u^0 \,,\\
u^{n+1}=\Lambda^{-1}\left[c_1u^{n}+c_2u^{n-1}+\dots +c_nu^{1}+b_nu^0\right] \,,
\end{array}
\right.
$$
where $\Lambda$ is given by (\ref{lainv}). The probability interpretation of the parameters 
$c_k$ and $b_n$ is still valid. In this case, however, we must use the {\it transpose matrix} $P$ 
of $A^{-1}$ to define the transition probabilities. 
Indeed, the $M\times M$ matrix $A^{-1}$ propagates in the positive semi-axis. Moreover, from (\ref{A}), 
it can be shown that $P$ defines a transition matrix (all the elements are positive numbers less than 
one and all the rows sum to one, see example below).

\vs \noindent
{\bf Example 4}
In the four dimensional case the matrix $P$ is
$$
P=
\left(
\begin{array}{cccc}
      1/(1+\mu) &   \mu/(1+\mu)^2 & \mu^2/(1+\mu)^3 & \mu^3/(1+\mu)^3\\
             0 &      1/(1+\mu) &   \mu/(1+\mu)^2 & \mu^2/(1+\mu)^2 \\
             0 &             0 &      1/(1+\mu) &     \mu/(1+\mu) \\
             0 &            0 &             0&             1 \\
\end{array}
\right)
\,.
$$

\vs
In the implicit case the random walk is defined as follows. Let the particle start in zero,
at the first step it could jump up to $M-1$ steps ahead with probabilities defined by the first $P$ row,
then we have the following partition of events:
\begin{itemize}
\item $E_{c_1}=$\Big{\{}{\it the particle starts in the previous position $x(t_{n})$ for instance $x_j$. Then, it could jump up to $M-j-1$ steps ahead with probabilities defined by the $j+1$-th $P$ row}\Big{\}};
\item $E_{c_k}=$\Big{\{}{\it the particle backs to the position $x(t_{n+1-k})$ for instance $x_j$. Then it could jump up to $M-j-1$ steps ahead with probabilities defined by the $j+1$-th $P$ row}\Big{\}};
\item $E_{b_n}=$\Big{\{}{\it the particle backs to the initial position $x(t_{0})$ then could jump up to $M-1$ steps ahead with probabilities defined by the first $P$ row}\Big{\}}.
\end{itemize}

\vs\noindent
{\bf Remark 13}
The implicit method is slower than the explicit one. 
However, we observe that, because of the stability constraint (\ref{stab}), 
the implicit scheme is advisable for small $\beta$.
Indeed, we have $\delta x\sim \delta t^\beta/\beta$, with $0<\beta<1$, and we are forced to raise the
time steps in order to improve ``spatial'' resolution.
\subsection{ggBm trajectories}
Along this section, we have provided a method to generate the random variable 
$L_\beta$ and the simulation results are shown in Figures 1-3. In particular
\begin{itemize}
\item
Figure 1 shows a random-walk  simulation with $\beta=0.4$.
In this case we used an implicit random-walk scheme. Moreover, we compared the histogram
evaluated over $N=10000$ simulations and the density function $M_{2/5}(x)$, $x\ge 0$.
\item
Figure 2 shows a random walk when $\beta=0.5$. Because $M_{1/2}(x)=\frac{1}{\sqrt{\pi}}e^{-x^2/4}$, $x\ge 0$, 
in this case $L_{1/2}=^{{}^{\!\!\!\!d}}|Z|$ where $Z$ is a  Gaussian random variable.
\item
Figure 3 shows a random walk with $\beta=0.8$. We used an
explicit random-walk scheme. Then, we compared the histogram with the
corresponding probability density function $M_{4/5}(x)$, $x\ge 0$.
\end{itemize}
\begin{figure}[t]
\begin{center}
\includegraphics[keepaspectratio=true,height=7cm]{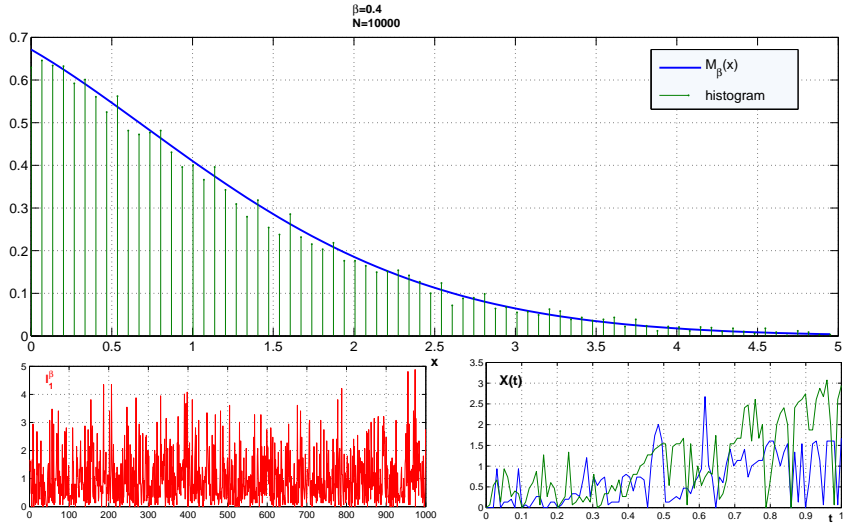}
\caption{
In the top panel, the histogram of $L_\beta$, which is calculated from a sample of $N=10000$ outcomes, is obtained
with an implicit random-walk scheme and it is compared with the exact PDF
$M_\beta(x)$, $x\ge 0$, with $\beta=0.4$.
In the bottom panels, the random variable $L_\beta$ (left) and
two trajectory examples (right) are shown.}
\end{center}
\label{figura 10}
\end{figure}
\begin{figure}[!t]
\begin{center}
\includegraphics[keepaspectratio=true,height=7cm]{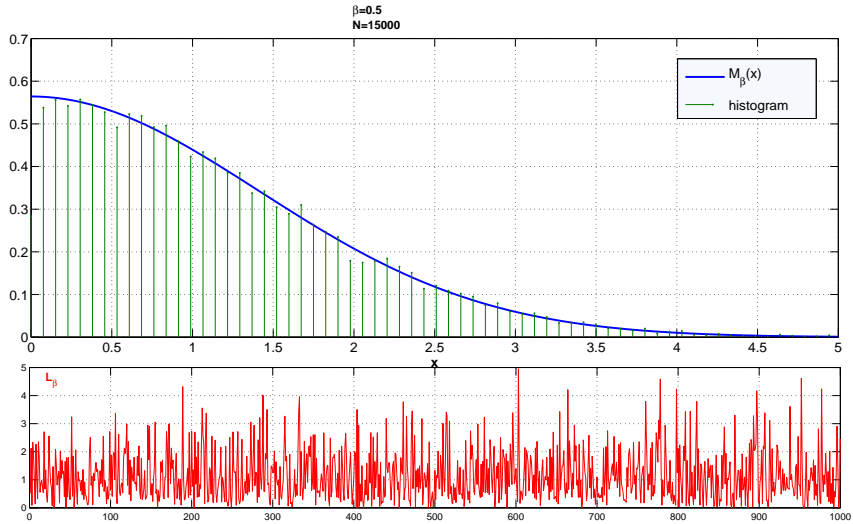}
\caption{
In the top panel, 
the histogram for the case $\beta=0.5$ is calculated from a sample of $N= 15000$ outcomes, 
which are obtained simulating independent Gaussian random variables. 
The corresponding $L_\beta$ is shown in the bottom.}
\end{center}
\label{figura 11}
\end{figure}
\begin{figure}[!t]
\begin{center}
\includegraphics[keepaspectratio=true,height=7cm]{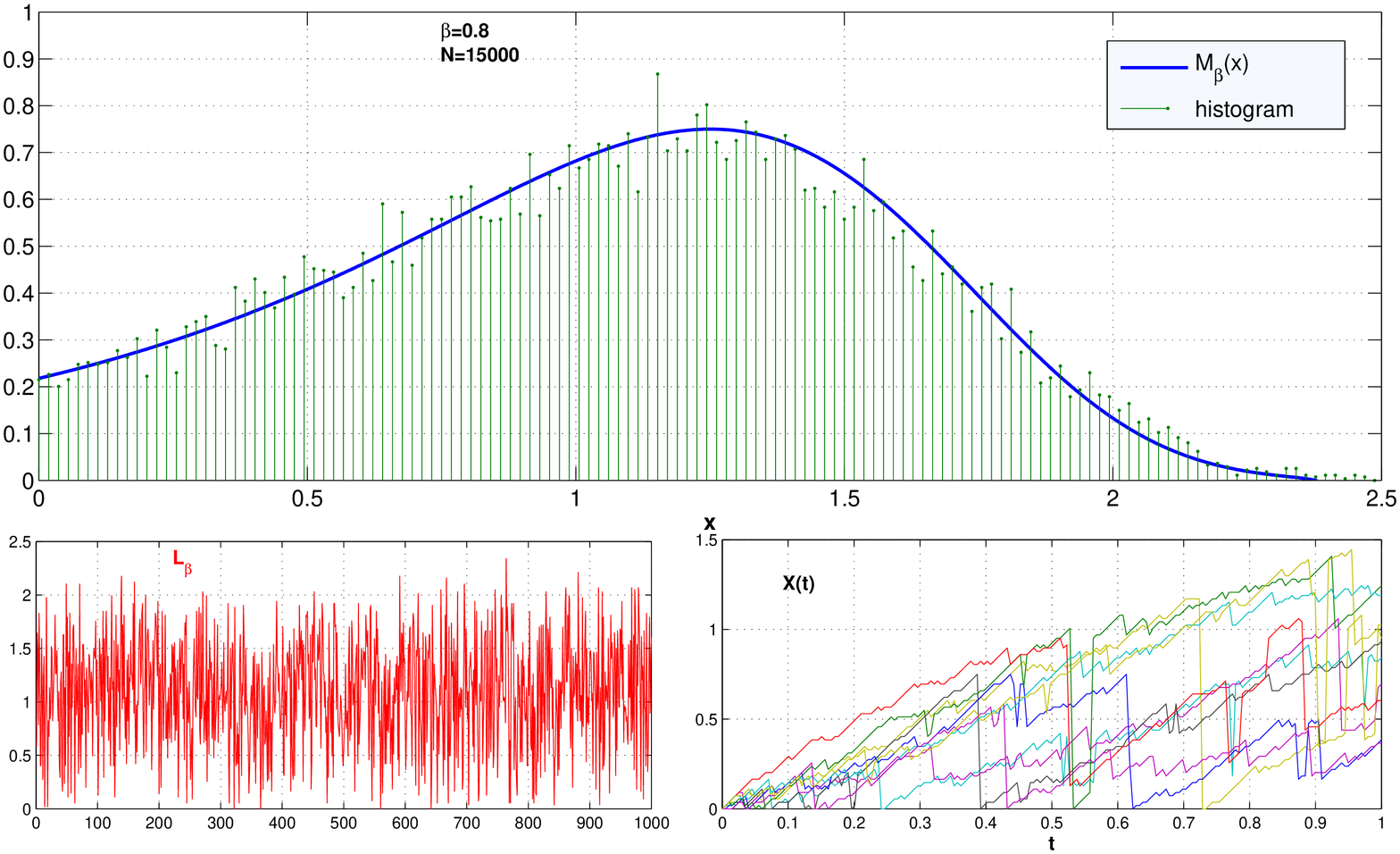}
\caption{
In the top panel, the histogram of $L_\beta$, which is calculated from a sample of $N=15000$ outcomes, 
it is  obtained with an explicit random-walk scheme and it is compared with the exact PDF
$M_\beta(x)$, $x\ge 0$, with $\beta=0.8$.
In the bottom panels, the random variable $L_\beta$ (left) and
many trajectory examples (right) are shown.}
\end{center}
\label{figura 12}
\end{figure}

\begin{figure}[!t]
\begin{center}
\includegraphics[keepaspectratio=true,height=7.5cm]{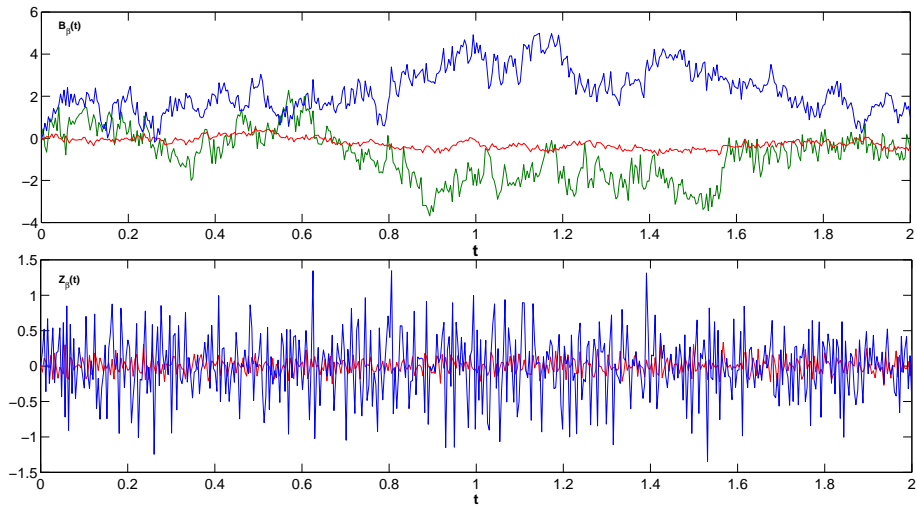}
\caption{$B_{\beta}(t)$ trajectories in the case $\beta=0.5$ (top panel) for $0\le t\le 2$. 
The time series
of the corresponding stationary noise $Z_\beta(t)$ is presented in the bottom panel.}
\end{center}
\label{figura 13}
\end{figure}
\begin{figure}[!h]
\begin{center}
\includegraphics[keepaspectratio=true,height=7.5cm]{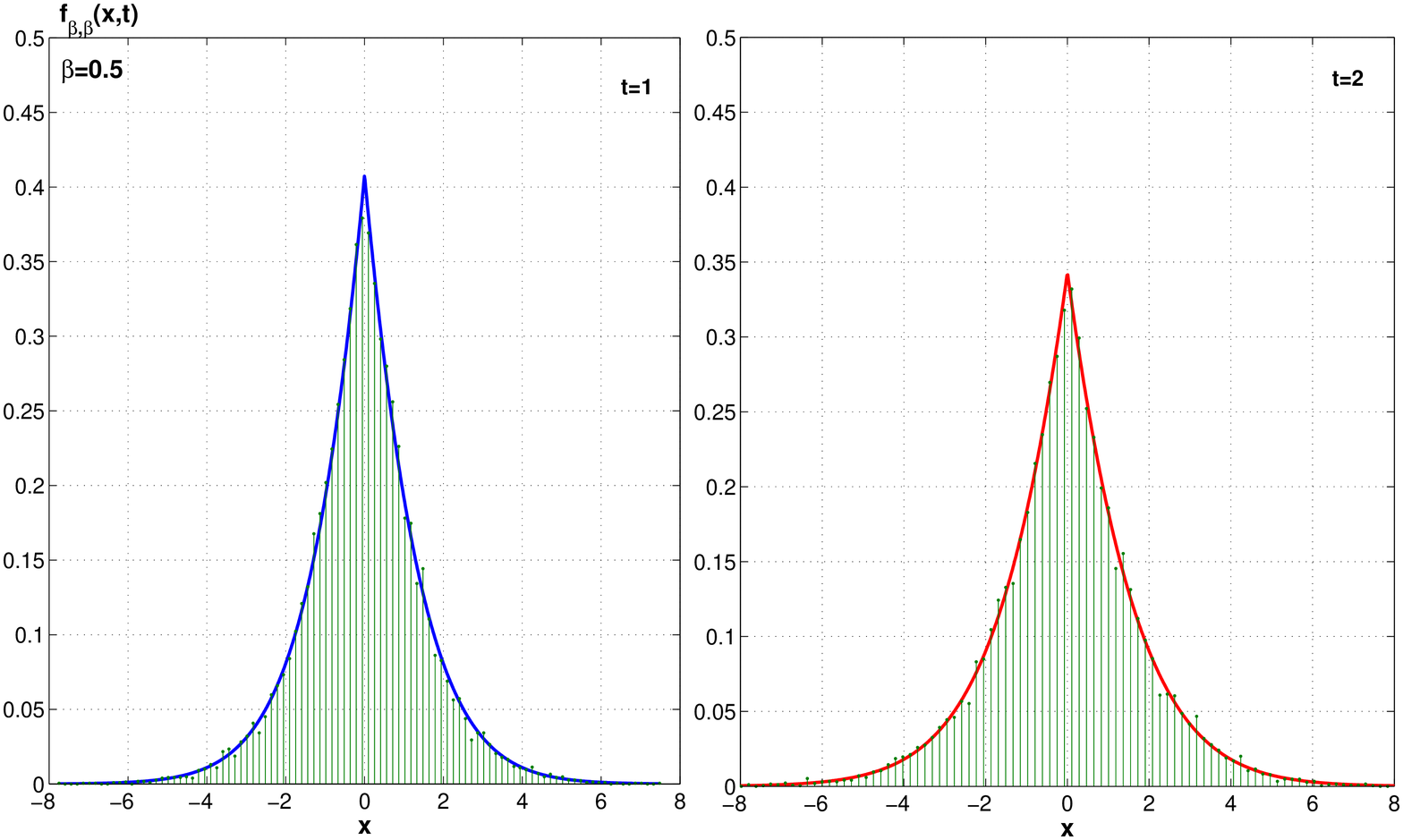}
\caption{Histograms of $N=15000$ simulations with $\beta=0.5$ and exact 
marginal density at $t=1$ and $t=2$.}
\end{center}
\label{figura 15}
\end{figure}

In all the studied cases, we have found a good agreement between the histograms and the theoretical 
density functions.\\

At this point, in order to obtain examples of the $B_{\alpha,\beta}(t)=\sqrt{L_\beta}X_\alpha(t)$ 
trajectories, we just
have to simulate the fractional Brownian motion $X_\alpha(t)$. For this purpose we have used an exact 
Cholesky method (see \cite{bardet}).\\

Path simulations of $B_{\beta,\beta}(t)$ (shortly $B_\beta$) and $B_{2-\beta,\beta}(t)$, with $\beta=1/2$ are shown
in Figures 4--9. 
The first process provides an example of stochastic model for slow-diffusion (short-memory),
the second provides a stochastic model for fast-diffusion (long-memory).
\begin{itemize}
\item Figure 4 shows some typical paths.
In the bottom panel, we present the corresponding increment process. Namely,
$$
Z_{\beta}(t_k)=B_{\beta}(t_{k})-B_{\beta}(t_{k-1}) \,, \quad t_k=k\delta t \,,
\quad k=1,2,\dots, M-1 \,.
$$
\item Figure 5 shows the agreement between simulations and the theoretical 
densities at times $t=1$ and $t=2$.
\item Figure 6 presents the plot of the sample variance in logarithmic scale. 
Moreover, we evaluated a
linear fitting, which shows a good agreement with the theoretical result.
\item Figure 7 shows some typical paths for the long-memory process
$B_{2-\beta,\beta}(t)$.
\item Figure 8 collects the histograms in the case $\alpha=2-\beta$ at time $t=1$ 
and time $t=2$.
The super-diffusive behavior (that is the rapid increasing of the variance in time) is highlighted.
\item Figure 9 presents the plot of the sample variance in logarithmic scale.
Even in this case, we evaluated a linear fitting.
\end{itemize}
\begin{figure}[!h]
\begin{center}
\includegraphics[keepaspectratio=true,height=7.5cm]{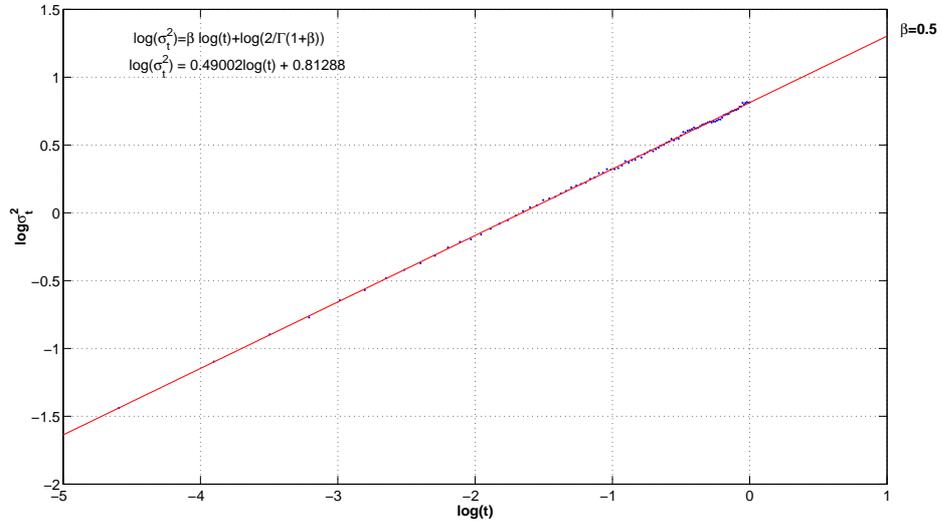}
\caption{Sample variance in logarithmic scale and linear fitting ($N=10^4$).}
\end{center}
\label{figura 16}
\end{figure}
\begin{figure}[!h]
\begin{center}
\includegraphics[keepaspectratio=true,height=7.5cm]{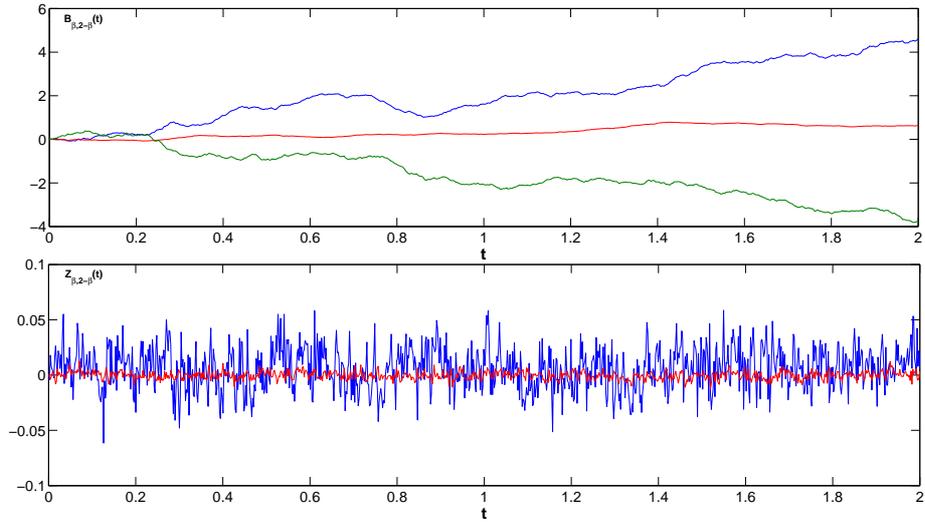}
\caption{$B_{2-\beta,\beta}(t)$ trajectories in the case $\beta=0.5$ (top panel) for $0\le t\le 2$. 
The corresponding stationary noise time series is presented in the bottom panel.}
\end{center}
\label{figura 17}
\end{figure}
\begin{figure}[!h]
\begin{center}
\includegraphics[keepaspectratio=true,height=7.5cm]{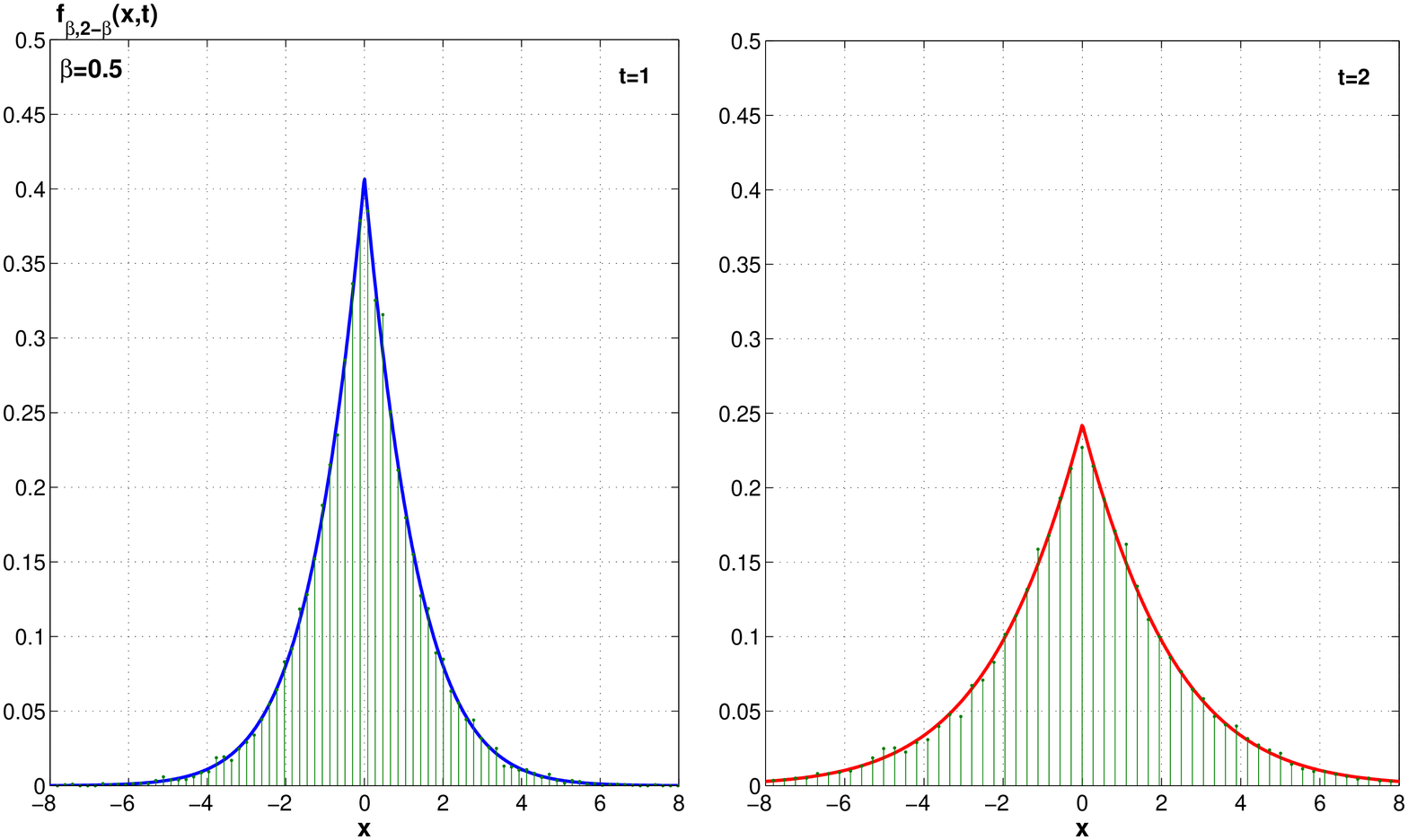}
\caption{Histograms of two 15000 running simulations of $B_{2-\beta,\beta}(t)$ with $\beta=0.5$ 
and exact distributions at different times $t=2$ and $t=1$.}
\end{center}
\label{figura 18}
\end{figure}
\begin{figure}[!h]
\begin{center}
\includegraphics[keepaspectratio=true,height=7.5cm]{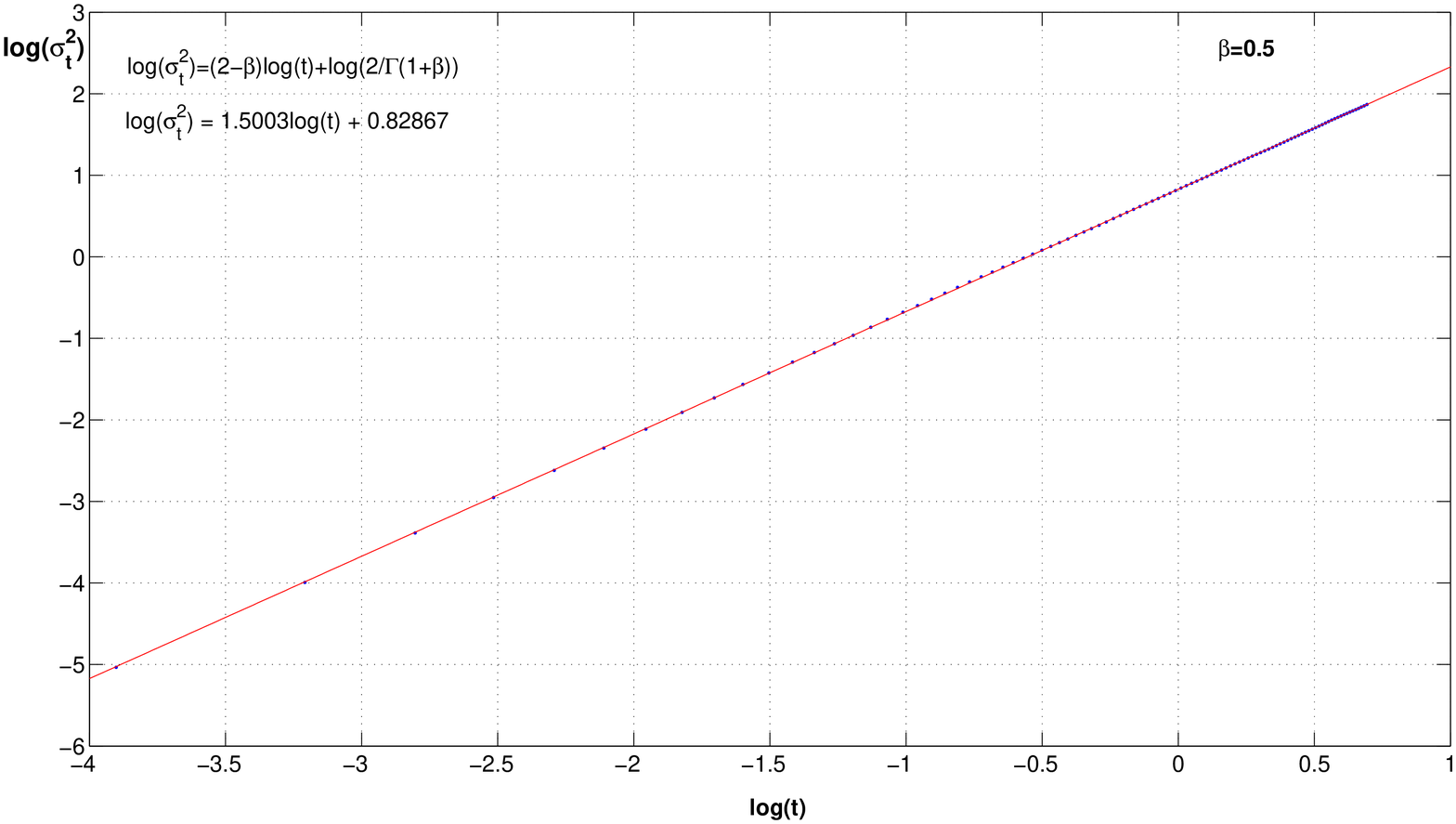}
\caption{Sample variance in logarithmic scale and linear fitting ($N=10^4$).}
\end{center}
\label{figura 19}
\end{figure}

\section{Concluding remarks}
The marginal probability density function (\ref{eqone})
of the generalized grey Brownian motion $B_{\alpha,\beta}(t)$, $t\ge 0$,
evolves in time according to a ``stretched'' time-fractional diffusion equation of
order $\beta$ (see \ref{geneq}). Therefore, the ggBm serves as stochastic model
for the anomalous diffusion described by these class of fractional equations.

The ggBm is defined canonically (see \ref{ggbmdef}) in the so called grey noise space 
$(\mathcal{S}'(\mathbb{R}), \mathcal{B}, \mu_{\alpha,\beta})$, where the grey noise measure 
satisfies (\ref{gnm}). 
However, the ggBm is an $H$-{\bf sssi} process of order $H=\alpha$ and Proposition \ref{pp1} 
provides a characterization of $B_{\alpha,\beta}(t)$ notwithstanding the underline probability space.

There are many other processes which serve as stochastic models for a given master equation.
In fact, given a master equation for a PDF $f(x,t)$,
it is always possible to define an equivalence class of
stochastic processes with the same marginal density function $f(x,t)$.
The ggBm defines a subclass $\{B_{\alpha,\beta}(t),\; t\ge 0\}$ 
associated to the non-Markovian equation (\ref{geneq}). 
In this case, the memory effects are enclosed in the typical dependence structure of a $H$-{\bf sssi} process. 
While, for instance in the case of a subordinated process 
(Examples 1 and 2), these are due to the memory properties of the {\it random time process}. 
The latter are preferable because provide a ready-made physical interpretation 
(Remark 3). 
However, $B_{\alpha, \beta}(t)$ is interesting because of the stationarity of its increments.

Proposition \ref{pp2} provides an enlighten representation of ggBm.
Thus, the generalized grey Brownian motion turns out to be merely a fractional Brownian
motion with stochastic variance, that is
$B_{\alpha,\beta}(t)=\Lambda_\beta X_\alpha(t)$, $t\ge 0$, where $\Lambda_\beta$ is a
suitable independent random variable (see \ref{ggrep}).
As a final remark, we observe that such a process is not ergodic,
as, heuristically, follows by the multiplication with the random variable
$\Lambda_\beta$. This appears also from the simulated trajectories. 
Indeed, it is impossible with a single realization of the system $B_{\alpha,\beta}(t,\omega)$, 
$\omega\in \Omega$, to distinguish a ggBm from a fBm with variance 
$2\Lambda_\beta^2(\omega)t^{\alpha}$, where $\Lambda_\beta(\omega)$ indicates a single 
realization of the random variable $\Lambda_\beta$.\\

As a further development of the present research, it is interesting to wonder if the 
generalized grey Brownian motion is the only one stationary increment process which serves 
as model for time-fractional diffusion equations like (\ref{geneq}).

\section*{Acknowledgements}
This work has been carried out in the framework of the research  project 
{\it Fractional Calculus Modelling} (URL: {\tt www.fracalmo.org}).
A. Mura appreciates partial support by the Italian Ministry of University (MIUR)
through the Research Commission of the  University of Bologna, 
and by the National Institute of Nuclear Physics (INFN) through the 
Bologna branch (Theoretical Group).
The authors are grateful to Professor F. Mainardi for his precious support in the 
preparation of this paper.

\section*{Appendix.  The $M$-function}\label{A1}
Let us define the function $M_{\beta}(z)$, $0<\beta<1$, as follows
\begin{eqnarray}
M_{\beta}(r) &=& \sum_{k=0}^{\infty}\frac{(-r)^k}{k!\Gamma\left[-\beta k+(1-\beta)\right]} 
\nonumber \\
&=& \frac{1}{\pi} 
\sum_{k=0}^{\infty}\frac{(-r)^k}{k!}\Gamma\left[(\beta(k+1)) \right]
\sin\left[\pi\beta(k+1)\right] \,, \quad r\ge 0 \,.
\label{functionM}
\end{eqnarray}
The above series defines a transcendental function (entire of order $1/(1 - \beta/2)$) 
of the Wright type,
introduced by Mainardi in \cite{MainardiAML96,MainardiCSF96},
(see also \cite{MainardiCISM97,Mainardi-Luchko-PagniniFCAA01}).
It is useful to recall some important properties of the $M$-function.
The best way to express these properties is in terms of the function
$$
\mathcal{M}_\beta(\tau,t)=t^{-\beta}M_\beta(\tau t^{-\beta}) \,, \quad \tau,t\ge 0 \,,
$$
which defines a probability density function in $\tau\ge 0$ for any $t\ge 0$ and $0<\beta <1$.
\begin{enumerate}
\item The following convolution formula holds:
\begin{equation}
\mathcal{M}_\nu(x,t)=\int_{0}^{\infty}\mathcal{M}_{\eta}(x,\tau)\mathcal{M}_{\beta}(\tau,t) \, d\tau \,,
\quad \nu=\eta\beta \,, 
\quad x\ge 0 \,,
\label{MW}
\end{equation}
where $\nu,\eta,\beta\in (0,1)$.
\item  The Laplace transform of $\mathcal{M}_\beta(\tau,t)$ with respect to $t$ is
\begin{equation}
\mathcal{L}\{\mathcal{M}_\beta(\tau,t);t,s\}=s^{\beta-1}e^{-\tau s^{\beta}} \,,
\quad \tau,s\ge 0 \,.
\label{pro1}
\end{equation}
\item  The Laplace transform of $\mathcal{M}_\beta(\tau,t)$ with respect to $\tau$ is
\begin{equation}
\mathcal{L}\{\mathcal{M}_\beta(\tau,t);\tau,s\}=E_\beta(-st^{\beta}) \,,
\quad t,s\ge 0\,,
\label{LaM}
\end{equation}
where $E_{\beta}(x)$ is the Mittag-Leffler function (\ref{equ1}).
\item The singular limit $\beta \rightarrow 1$ gives
\begin{equation}
\mathcal{M}_1(\tau,t)=\delta(\tau-t) \,, \quad \tau,t\ge 0 \,.
\label{pro2}
\end{equation}
\item Let $G(x,t)$ be the Gaussian density function, then
\begin{equation}
G(x,t)=\frac{1}{2}\mathcal{M}_{1/2}(|x|,t)=
\frac{1}{\sqrt{4\pi t}} \exp\left(-\frac{x^2}{4t}\right) \,.
\label{GM}
\end{equation}
\end{enumerate}


\end{document}